\documentclass[aps,prb,twocolumn,superscriptaddress,10pt]{revtex4-2}

\usepackage{graphicx}
\usepackage{epstopdf}
\usepackage{amsmath, amsfonts, amssymb}
\usepackage{physics}
\usepackage{mathtools}
\usepackage{siunitx}
\usepackage[colorlinks=true,citecolor=blue,linkcolor=blue,urlcolor=blue]{hyperref}
\usepackage{xcolor}
\usepackage{ctex}

\usepackage[normalem]{ulem}

\newcommand{\corr}[1]{#1}

\makeatletter
\renewcommand{\today}{\ifcase\month\or
  January\or February\or March\or April\or May\or June\or
  July\or August\or September\or October\or November\or December\fi
  \space\number\day, \number\year}
\makeatletter

\begin{document}

\title{Quantum Boomerang Effect in Time-Crystalline Structures}

\author{Qi-wen Peng}
 \affiliation{School of Physics and Optoelectronics, Xiangtan University, Hunan 411105, China}
 \author{Krzysztof Sacha}%
\affiliation{
Instytut Fizyki Teoretycznej, Uniwersytet Jagiello\'nski, ulica Profesora Stanis\l{}awa \L{}ojasiewicza 11, PL-30-348 Krak\'ow, Poland}
\author{Chu-hui Fan}
 \affiliation{School of Physics and Optoelectronics, Xiangtan University, Hunan 411105, China}

\date{March 18, 2026}

\renewcommand{\abstractname}{Abstract} 
\begin{abstract}
The quantum boomerang effect (QBE) is a unique dynamical signature of Anderson localization, characterized by a launched wavepacket that initially drifts but ultimately returns to its initial position due to fundamental quantum interference. In this work, we theoretically establish and quantitatively characterize the QBE in a time-crystalline structure using a periodically driven quantum particle in a one-dimensional potential well. By constructing maximally localized Floquet-Wannier states and introducing temporal disorder, we rigorously map the continuous Floquet dynamics onto a discrete disordered tight-binding lattice. By positioning a detector at a fixed spatial coordinate, we monitor the temporal evolution of the wavepacket, to extract the mean temporal center of mass of the probability density in a time-crystalline structure. This mean temporal center of mass exhibits an initial ballistic expansion, followed by a pronounced U-turn, and ultimately returns to its initial temporal position after long-time evolution. These results confirm the existence of the complete QBE in the time domain. They also demonstrate that non-trivial dynamics can be explored within time-crystalline systems, even though these structures already possess an inherent temporal periodicity.
\end{abstract}

\maketitle
\section{Introduction}

Anderson localization---the suppression of wave diffusion by disorder-induced quantum interference~\cite{Anderson1958,RevModPhys.80.1355}---is traditionally characterized by the exponential spatial localization of eigenstates. Recently, a distinct dynamical signature of localization was identified: the quantum boomerang effect (QBE)~\cite{PhysRevA.99.023629}. When a quantum wavepacket is launched with a finite mean velocity in a disordered medium, it does not relax to a halted state over a transport mean free path, as dictated by classical diffusion. Instead, the disorder-averaged center of mass exhibits an initial ballistic drift, undergoes a reversal of motion, and returns to its initial position at long times~\cite{PhysRevA.99.023629,PhysRevA.103.063316}. This phenomenon originates from the intricate quantum interference mechanisms that drive Anderson localization~\cite{PhysRevA.99.023629,PhysRevA.103.063316}, providing a clear dynamical distinction from classical transport~\cite{PhysRevE.111.024104}. Since its initial theoretical proposal, the QBE has been experimentally confirmed in the quantum kicked rotor~\cite{PhysRevX.12.011035} and disordered photonic lattices~\cite{Hou2026}. Theoretically, it has been rapidly extended to a diverse array of complex scenarios, including interacting many-body gases~\cite{PhysRevA.102.013303,PhysRevB.107.094204}, non-Hermitian systems~\cite{PhysRevB.106.104310}, and various generalized lattice models subject to synthetic gauge fields or broken symmetries~\cite{PhysRevB.105.L180202,PhysRevB.106.L060301}. These studies establish the QBE as a robust and universal dynamical feature of transport breakdown.

Analogous to the spontaneous formation of spatial crystals, time crystals can emerge spontaneously in periodically driven many-body systems \cite{PhysRevLett.109.160401,PhysRevA.91.033617,PhysRevLett.117.090402,PhysRevLett.116.250401,Zhang2017,Choi2017,Sacha_2018,Sacha2020_Book,zaletelColloquiumQuantumClassical2023}. Furthermore, condensed matter phases can be investigated within time-crystalline structures (or phase-space crystalline structures \cite{guoPhaseSpaceCrystals2013,guoBook2021}) engineered through external periodic perturbations \cite{Sacha2015,Sacha2020_Book,giergielTimetronicsTemporalPrinted2025}. This temporal periodicity enables the construction of a temporal Wannier basis, thereby allowing the effective Hamiltonian to be mapped onto a synthetic temporal lattice~\cite{PhysRev.138.B979,Sacha2020_Book}. Introducing temporal disorder into such systems leads to Anderson localization in the time domain~\cite{Sacha2015,PhysRevA.94.023633,Sacha2020_Book,PhysRevB.111.L180201}. Because Anderson localization is the fundamental prerequisite for the standard QBE~\cite{PhysRevA.99.023629,PhysRevA.103.063316}, its emergence in the time domain naturally poses a question regarding the temporal QBE: whether a wave packet, starting from an initial temporal coordinate, will ultimately return to its temporal point of origin.

In this paper, we theoretically establish and quantitatively characterize the quantum boomerang effect entirely within the time domain, utilizing a periodically driven quantum particle in a one-dimensional potential well as our physical platform. Time-dependent processes are inherently natural in spatial crystals, as time serves as an additional dimension that allows us to track the evolution of a system's spatial properties. In time-crystalline structures, however, regular behavior already occurs within the temporal domain itself. Thus, it may seem that there is no room for further non-trivial dynamics, as the temporal degree of freedom appears already 'occupied.'
In periodically driven systems, a time-crystalline framework consists of a finite set of temporal unit cells that repeat periodically. This establishes a fundamental time scale associated with the periodicity of the crystalline order. We observe this order by fixing a detector at a specific spatial position and analyzing the detection probability over time. A change in this probability across successive periods indicates a non-trivial evolution of the system state within the time-crystalline lattice. To visualize this evolution, one must analyze the probability density as a function of time modulo the repetition period of the underlying crystalline structure.
In this manner, we realize and describe the QBE in the time domain. Our observations reveal that the wave packet is initially concentrated at progressively later moments within each repetition period. Subsequently, the probability density begins to shift back toward earlier times, ultimately returning to the same temporal position it occupied at the start of the evolution.

The paper is organized as follows: Section~\ref{sec:model} introduces the exact model of the driven particle, derives the effective secular Hamiltonian, and details the construction of Floquet-Wannier states and mapping of the effective Hamiltonian onto a tight-binding temporal lattice. Section~\ref{sec:dynamics} presents the numerical evolution, analyzing the disorder-driven temporal quantum boomerang dynamics. Finally, Section~\ref{sec:discussion} summarizes our findings. 

\begin{figure}[t]
\centering
\includegraphics[width=0.9\linewidth, clip]{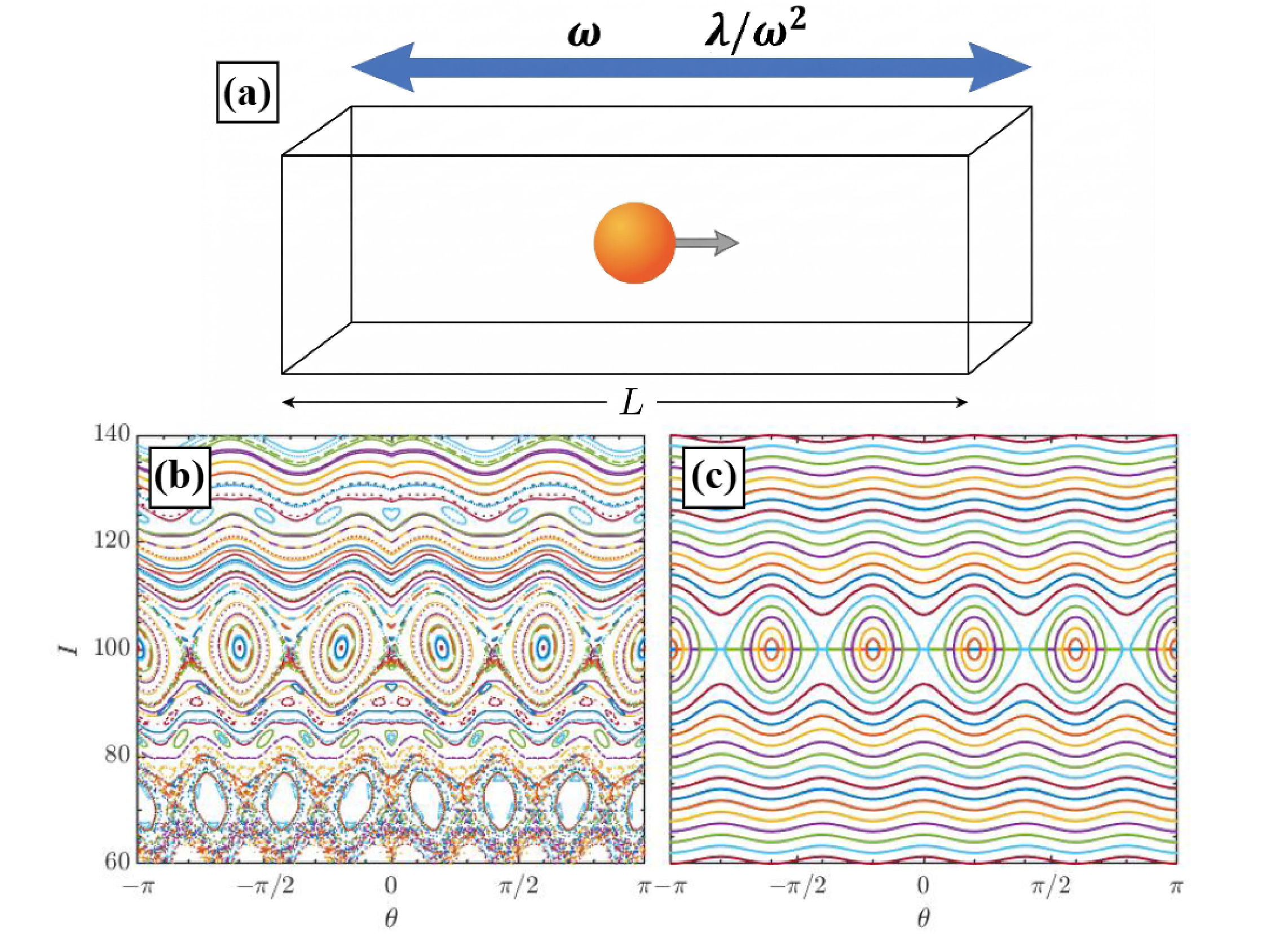}
\caption{(a) A particle moves freely within a box of length $L$, undergoing perfectly elastic reflections at the boundaries. The entire box is periodically shaken with frequency $\omega$ and amplitude $\lambda/\omega^2$, as indicated by the double-headed arrow above the box. The small arrow on the particle represents its initial momentum. (b) The exact stroboscopic phase-space portrait of the system governed by the exact Hamiltonian~(\ref{eq:exact_ham}). (c) The corresponding phase-space structure generated by the effective Hamiltonian~(\ref{eq:final_eff_ham}) under the secular approximation for $s=5$. The system parameters used in the simulations are set to $L=\pi$, $\omega=500$, and $\lambda=1000$.}
\label{fig:schematic_model}
\end{figure}

\section{Model} \label{sec:model}

As illustrated in Fig.~\ref{fig:schematic_model}, here we consider a particle of unit mass moving in a box that is periodically shaken in time. Let us first assume that the shaking is harmonic in time with frequency $\omega$ and amplitude $\lambda/\omega^2$. In the frame moving with the box, this shaking results in a time-oscillating linear potential, and the Hamiltonian of the system is given by:
\begin{equation}
\label{eq:exact_ham}
H_0=\frac{p^{2}}{2}+\lambda x\cos(\omega t).
\end{equation}
We assume the box size to be $L=\pi$ and the frequency $\omega$ to satisfy the resonance condition with the frequency, $\Omega=|p|$, of the classical unperturbed motion of the particle, i.e., $\omega=s\Omega$, where $s$ is an integer.

Let us start with a classical analysis of the resonant motion and then switch to the quantum description. It is convenient to describe the classical particle behavior with the action-angle variables \cite{Lichtenberg1992}, i.e., $\theta\in[-\pi,\pi)$ where $x=|\theta|$, $I=\lvert p \rvert$ and the Hamiltonian, 
\begin{equation}
\label{H0_action}
    H_0=\frac{I^2}{2}+\lambda\cos(\omega t)\sum_{n=-\infty}^{\infty}h_n e^{in\theta},
\end{equation}
where $h_0=\pi/2$, $h_n=-2/(\pi n^2)$ for odd $n$, and $h_n=0$ for even $n$.

To understand the effective dynamics, we switch to the frame moving along the resonant trajectory via the canonical transformation, $\tilde{I} = I$ and $\tilde{\theta}=\theta-\omega t/s$, where $s$ is an odd integer. For $\tilde I$ close to the resonant value $I_s=\omega/s$, within the secular approximation \cite{Lichtenberg1992}, the Hamiltonian can be averaged over time, yielding the effective Hamiltonian \corr{(see Appendix~A)}:
\begin{equation}
\label{H0eff1}
\corr{H_{0,\mathrm{eff}}^{C}} = \frac{\tilde I^2}{2} - \frac{\omega}{s}\tilde{I} -\lambda\frac{2}{\pi s^{2}}\cos(s\tilde{\theta}).
\end{equation}
To investigate the behavior close to the resonant orbit, we perform a Taylor expansion of the Hamiltonian (\ref{H0eff1}) around the resonant action $I_s$, which leads to the final effective Hamiltonian:
\begin{equation}
\label{eq:final_eff_ham}
\corr{H_{0,\text{eff}}^{C}} \approx \frac{P^{2}}{2} + \tilde{V}\cos(s\tilde{\theta}).
\end{equation}
This result indicates that, if we switch to the frame moving along the resonant trajectory, a particle resonantly driven in a periodically shaken box can be approximately mapped to a particle in a static lattice with the effective momentum defined as $P=\tilde{I}-I_s$ and the amplitude of the effective lattice potential $\tilde{V}= -\frac{2\lambda}{\pi s^2}$. Consequently, for $s \gg 1$, we are dealing with a particle moving in a time-independent crystalline potential. In Fig.~\ref{fig:schematic_model}, we present a comparison between the exact stroboscopic phase-space portrait and the structure generated by the effective Hamiltonian, which illustrates the validity of the secular approximation for the system parameters chosen in this work.

\corr{It is important to emphasize that the secular approximation is not expected to reproduce the full phase-space portrait of the exact driven system, but to capture the dynamics near the resonant action $I_s$. Note that both descriptions exhibit the same dominant $s=5$ resonant islands in the vicinity of $I_s=100$, which illustrates the validity of the effective Hamiltonian for constructing the time-crystalline lattice. The remaining fine structures in Fig.~\ref{fig:schematic_model}(b) arise from non-resonant and higher-order oscillating terms, which may induce micromotion or small higher-order corrections but do not affect the leading-order resonant dynamics. Therefore, the effective Hamiltonian should be understood as a local description associated with the chosen $s$-th resonance. If the initial conditions are chosen far away from $I_s$, a corresponding effective Hamiltonian should be derived.}

\corr{To proceed with the quantum description, a quantum version of the secular approximation can be applied on Eq.~(\ref{eq:exact_ham}) \cite{Berman1977} (see Appendix~B).} While the classical description provides physical intuition regarding the effective behavior of the system, it does not allow for a straightforward derivation of the wavefunction in the original Cartesian coordinates, as the canonical transformation to action-angle variables is nonlinear. As will be shown shortly, the quantum description bypasses this issue, however, it yields an effective Hamiltonian in matrix form, which obscures the expected qualitative behavior of the system. By combining both methodologies, we leverage the physical intuition gained from the classical approach while maintaining a rigorous quantitative description provided by the fully quantum treatment.

To investigate the particle behavior in the quantum description, we first introduce the unitary transformation $U(t) = \exp(-i \hat{n} \frac{\omega}{s} t)$, where $\hat{n} \psi_n=n\psi_n$ with $\psi_n=\sqrt{2/\pi}\sin(nx)$ being the eigenbasis of the particle in the unperturbed box potential. This transformation corresponds to the classical transformation to the frame moving along the resonant trajectory. In the moving frame, averaging the Hamiltonian over time \cite{Berman1977}, we obtain the following matrix elements of the \corr{quantum} effective Hamiltonian:
\begin{equation}
\label{effective_matrix}
\begin{aligned}
\langle n' | \corr{H_{0,\text{eff}}^{Q}}| n \rangle = &\left( \frac{n^2}{2} - \frac{\omega}{s} n \right) \delta_{n',n} \\
&- \frac{\lambda}{\pi} \left[ \frac{1}{s^2} - \frac{1}{(2n+s)^2} \right] \delta_{n',n+s} \\
&- \frac{\lambda}{\pi} \left[ \frac{1}{s^2} - \frac{1}{(2n-s)^2} \right] \delta_{n',n-s}.
\end{aligned}
\end{equation}
The eigenstates of this Hamiltonian correspond to the Floquet states \cite{PhysRev.138.B979} of the original problem (\ref{eq:exact_ham}) within the secular approximation --- they evolve with a period of $T=2\pi/\omega$ when transformed back to the laboratory frame.

For large $s$, the eigenvalues of the Hamiltonian (\ref{effective_matrix}) organize into energy bands. By restricting our analysis to the first band, we can construct Wannier states, $w_j$, localized at the sites of the effective lattice potential [cf. Eq.~(\ref{eq:final_eff_ham})] through a suitable superposition of these eigenstates. In the Wannier basis, the effective Hamiltonian restricted to the first band can be expressed as a tight-binding model:
\begin{equation}
\label{tb}
\corr{H_{0,\text{eff}}^{Q}} = -J \sum_{j=1}^s \left( |w_j\rangle \langle w_{j+1}| + |w_{j+1}\rangle \langle w_j| \right),
\end{equation}
where the tunneling amplitude $J = -\langle w_j | \corr{H_{0,\text{eff}}^{Q}} | w_{j+1} \rangle$ can be controlled by tuning the parameters $\lambda$ and $\omega$ and an irrelevant constant term has been omitted.

\begin{figure}[tb]  
\centering
\includegraphics[width=\columnwidth]{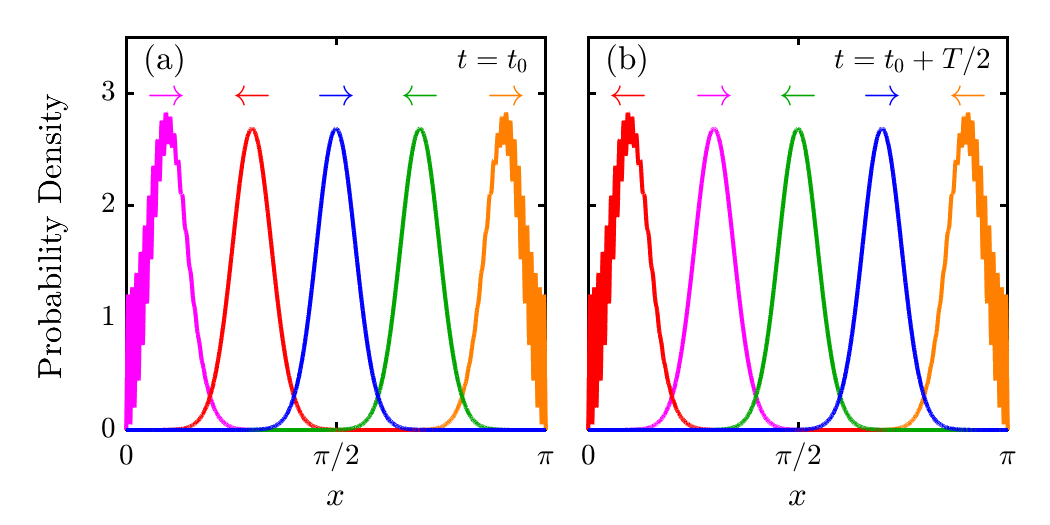} 
\caption{Probability densities of the $s=5$ Floquet-Wannier states as a function of the position $x$ at two different moments in time: (a) $t = t_0$ (with $t_0 = T/4$) and (b) $t = t_0 + T/2$. Different colors are used to distinguish the distinct Wannier states, and the arrows over the wave packets indicate their instantaneous direction of motion at the corresponding moments. The system parameters are same as those in Fig.~\ref{fig:schematic_model}.}
\label{fig:wannier}
\end{figure}

When observed in the laboratory frame, the Wannier states $w_j(x,t)$ evolve periodically in time, as illustrated in Fig.~\ref{fig:wannier} --- hereafter, we shall refer to them as Floquet-Wannier states. For instance, the ground state of the Hamiltonian (\ref{effective_matrix}) is a uniform superposition of all Floquet-Wannier states. Consequently, if a detector is placed near a wall of the box potential in the laboratory frame, the detection probability will oscillate periodically. In this case, the entire time-crystalline structure (comprising $s$ temporal sites) repeats with a period of
\begin{equation}
\label{Ts}
    T_s=sT=s\frac{2\pi}{\omega},
\end{equation}
which corresponds to the period of the resonant trajectory.

To introduce disorder into the tight-binding model (\ref{tb}), we apply an additional modulation to the shaking of the box. Specifically, we consider a particle described by the Hamiltonian
\begin{equation}
H = H_0 + x f(t),
\end{equation}
with 
\begin{equation}
\label{f_of_t}
f(t) = \frac{\gamma}{\sqrt{s+1}} \sum_{k=-s}^s \frac{e^{-i( k \omega t / s+\phi_k)}}{h_k},
\end{equation}
where the sum runs over odd values of $k$, $h_k$ are defined in (\ref{H0_action}), and $\phi_k=-\phi_{-k}$ are random phases uniformly distributed over $[0, 2\pi)$. Within the secular approximation, this leads to an additional term in the effective Hamiltonian compared to (\ref{eq:final_eff_ham}):
\begin{equation}
\corr{H_{\text{eff}}^{C}} \approx \frac{P^{2}}{2} + \tilde{V}\cos(s\tilde{\theta}) + \frac{\gamma}{\sqrt{s+1}} \sum_{k=-s}^s e^{-i(k\tilde{\theta} + \phi_k)}.
\end{equation}
In the tight-binding approximation, the \corr{quantum} effective Hamiltonian takes the form:
\begin{equation}
\label{tb_full}
\corr{H_{\text{eff}}^{Q}} = -J \sum_{j=1}^s \left( |w_j\rangle \langle w_{j+1}| + |w_{j+1}\rangle \langle w_j| \right) + \sum_{j=1}^s \epsilon_j |w_{j}\rangle \langle w_j|,
\end{equation}
where the on-site energies $\epsilon_j$ are given by:
\begin{eqnarray}
\epsilon_j &=& \frac{\gamma}{\sqrt{s+1}} \sum_{k=-s}^s \int_0^{2\pi} d\tilde{\theta}\; |w_j(\tilde{\theta})|^2\; e^{-i(k\tilde{\theta} + \phi_k)} \cr
&\approx& \frac{2\gamma}{\sqrt{s+1}} \sum_{k=1}^s \cos(k\tilde{\theta}_j + \phi_k).
\end{eqnarray}
Here, we assume that the probability densities of the Wannier functions $|w_j(\tilde{\theta})|^2$ can be approximated by Dirac delta functions centered at $\tilde{\theta}_j=(2j-1)\pi/s$. By invoking the central limit theorem, it follows that for large $s$, the $\epsilon_j$ values behave as independent random variables following a Gaussian distribution with zero mean and standard deviation $\gamma$.

In Fig.~\ref{fig:wannier}, we present the Floquet-Wannier states obtained for a box of size $L=\pi$, where the driving frequency $\omega$ is $s=5$ times larger than the frequency of the resonant trajectory. This represents a small time-crystalline structure, as it contains only five temporal sites. However, these Floquet-Wannier states can be utilized to describe a significantly larger system if allowed to evolve over a much broader range. For instance, by increasing the box size fifty-fold, we obtain a time-crystalline structure comprising $s=250$ temporal unit cells. Remarkably, this scaling is achieved without the need to re-evaluate the Floquet-Wannier states or the parameters of the tight-binding model (\ref{tb_full}).

\section{Temporal Quantum Boomerang Effect} \label{sec:dynamics}

\begin{figure}[tb]
\centering
\includegraphics[width=\columnwidth]{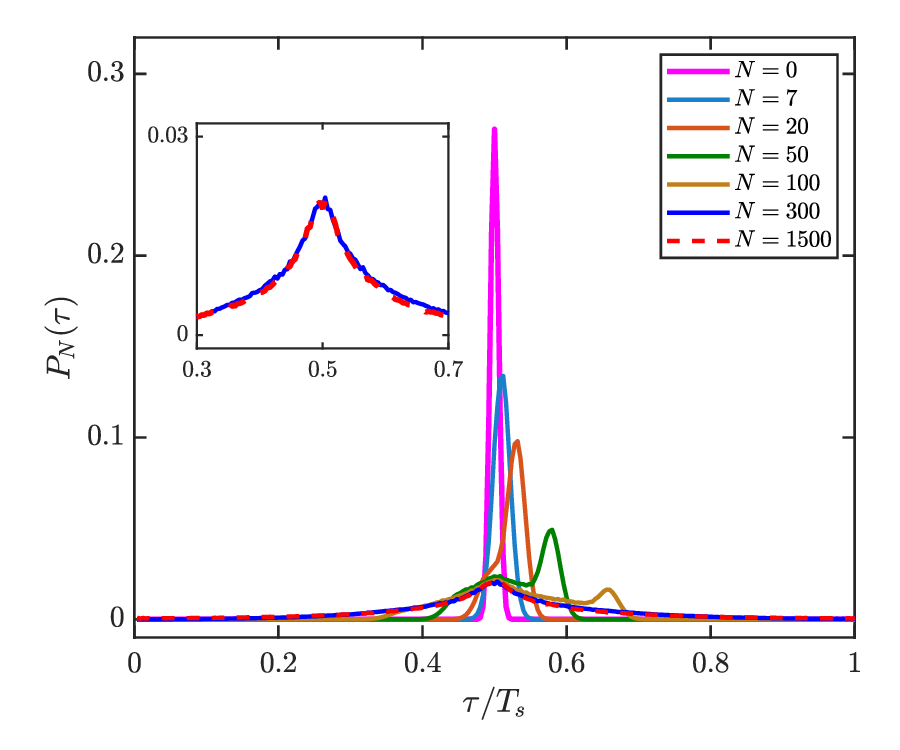} 
\caption{Probability density $P_N(\tau)$ for the detection of the particle at position $x_0=0.5\pi$ in the box, shown as a function of \corr{the scaled temporal coordinate $\tau/T_s$. The horizontal axis covers one resonant period, corresponding to one period of the time-crystalline structure.} The time-crystalline structure extends over $s=250$ driving periods $T$. The probability density is plotted for various values of $N$, i.e., after a different number of periods $T_s=sT$ of the resonant particle trajectory. For small $N$, the density $P_N(\tau)$ reveals a drift toward later time moments $\tau$, followed by a return to the initial temporal location, demonstrating the QBE in the time domain. The inset shows a magnified comparison of $P_N(\tau)$ for $N=300$ (blue solid line) and $N=1500$ (red dashed line). All curves are averaged over $2000$ realizations of temporal disorder. System parameters are the same as in Fig.~\ref{fig:schematic_model}. The resulting nearest-neighbor hopping amplitude in the tight-binding model (\ref{tb_full}) is $J \approx 6.63\times10^{-2}$. The initial Gaussian wave packet has a location $j_0=125$, a width of $\sigma = 4$ and a momentum of $k_0 = \pi/2$ [see Eq.~(\ref{gaussian})]. The disorder strength is $\gamma=0.5J$.}
\label{fig:density_W1}
\end{figure}

To investigate the QBE dynamics, we consider a time-crystalline structure comprising $s=250$ temporal lattice sites which, within the secular approximation, is described by the tight-binding model (\ref{tb_full}). Temporal disorder is created by a superposition of $s$ harmonics with randomly chosen phases (\ref{f_of_t}) which results in random on-site energies $\epsilon_j$ described by the normal distribution with variance $\gamma^2$. We expand the state of the particle in the Floquet-Wannier state basis, $\Psi(x,t)=\sum_j c_j(t) w_j(x,t)$, where the coefficients $c_j$ obey initially a Gaussian distribution:
\begin{equation}
\label{gaussian}
c_j(t=0) = \mathcal{N} \exp\left[ -\frac{(j-j_0)^2}{2\sigma^2} \right] e^{i k_0 j}.
\end{equation}
Here $\mathcal{N}$ is the normalization factor, while $\sigma$ and $j_0$ denote the width and the center of the initial wave packet, respectively. The index $j$ labels the $j$th Wannier-Floquet wave packet, i.e. the $j$th temporal site of the time-crystalline structure, and $k_0$ determines the initial group velocity $v_0=2J\sin(k_0)$~\cite{PhysRevA.103.063316}. 

Now we can analyze the QBE in the time domain. In Fig.~\ref{fig:density_W1} we present the temporal evolution of the probability density for the detection of the particle at a fixed position $x_0$ \corr{(close to the wall of the potential box)} in the laboratory frame,
\begin{equation}
\label{temporal_density}
    P_N(\tau)=\frac{\lvert \Psi (x_0,NT_s+\tau) \rvert ^2}{\int _{0} ^{T_s} \;\lvert \Psi (x_0,NT_s+\tau) \rvert ^2 \;d \tau},
\end{equation}
\corr{where $N$ denotes the index of the resonant-trajectory period. To clarify the dual role of time in our framework, we decompose the laboratory time as $t=N T_s+\tau$. Here, $N T_s$ is the slow evolution time, analogous to the observation time in the conventional spatial quantum boomerang effect, while $\tau=t \mod T_s$ is the temporal coordinate within one period of the time-crystalline structure. In this sense, the temporal quantum boomerang effect describes the drift and return of the probability distribution along the temporal coordinate $\tau$ as the slow evolution time $N T_s$ increases. The probability density $P_N(\tau)$ therefore provides a time-resolved picture of the temporal displacement within the period of the time-crystalline lattice.}

It is interesting to note that during the first several periods, $T_s$, the wave packet reveals gradual spreading accompanied by a shift of its peak toward later times. However, after long-time evolution, i.e. after about $300T_s$, the peak of the wave packet returns to its initial time, while the spreading of the wave packet becomes frozen.

These features are characteristic signatures of the QBE and Anderson localization, but now manifested in the time domain. \corr{To demonstrate the temporal QBE experimentally, repeated measurements of the particle position at a given time $t$ should be performed, which allows for the reconstruction of the probability density $|\psi(x,t)|^2$. If $|\psi(x,t)|^2$ is known for any $x$ and $t$, the temporal QBE can be demonstrated by choosing $x=x_0$ close to the wall of the potential box and plotting normalized $|\psi(x_0,t)|^2$, Eq.~\eqref{temporal_density}, in the interval $t\in[0,T_s)$, and subsequently in successive intervals $t\in[NT_s,(N+1)T_s)$. In this way, we can reproduce the results presented in Fig.~\ref{fig:density_W1}. Specifically, they demonstrate that with the highest probability, the particle initially appears progressively later within the period $T_s$, but over the course of time, the particle begins to appear progressively earlier, and finally, its appearances return to the initial moment within this period.}

\begin{figure}[tb]
\centering
\includegraphics[width=\columnwidth]{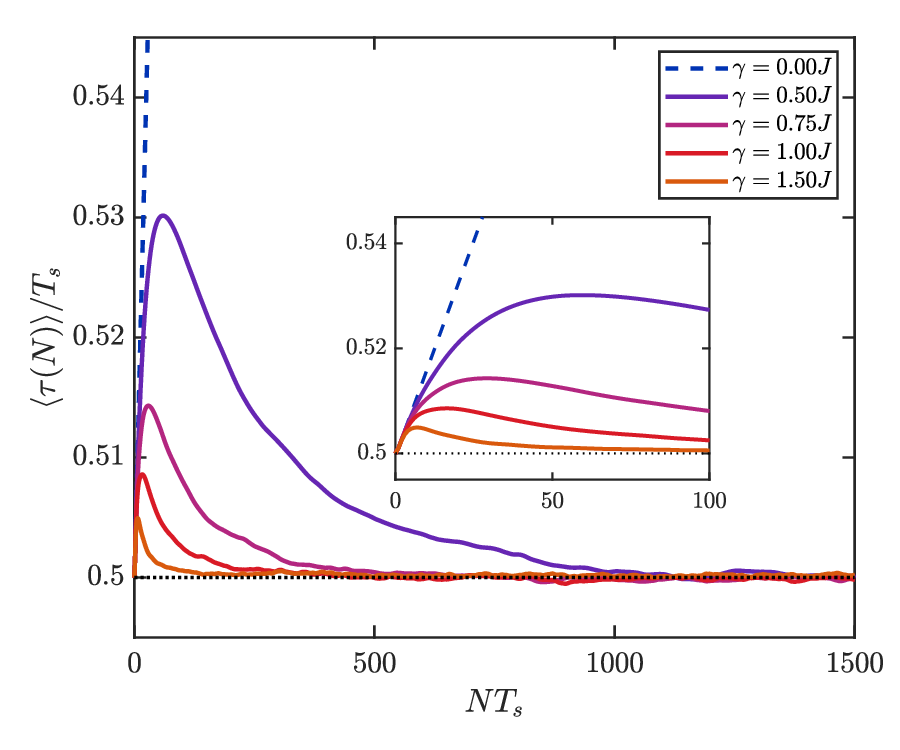} 
\caption{\corr{Temporal center of mass $\langle\tau(N)\rangle/T_s$ as a function of the slow evolution time $NT_s$, see Eq.~(\ref{t-mass}). The vertical axis is normalized by the resonant period $T_s$.} The figure displays the dynamics of the temporal center of mass for various disorder strengths $\gamma$, as indicated in the legend. The inset focuses on small values of $NT_s$. The horizontal black dotted line indicates the initial temporal center of mass, and all other parameters are the same as in Fig.~\ref{fig:density_W1}.}
\label{fig:centroid_trajectories}
\end{figure}

In Fig.~\ref{fig:centroid_trajectories}, we further quantify the temporal QBE dynamics by introducing the temporal center of mass, defined as:
\begin{equation}
\langle \tau(N) \rangle = \int _{0} ^{T_s} P_N(\tau) \; \tau \; d \tau.
\label{t-mass}
\end{equation}
This quantity characterizes the average temporal position of the particle in the time-crystalline structure and provides a convenient way to track its drift-and-return dynamics in the time domain. As shown in Fig.~\ref{fig:centroid_trajectories}, in the absence of temporal disorder ($\gamma=0$), the temporal center of mass moves toward later and later times and does not show any evident return within the considered time window. Once temporal disorder is introduced, however, a clear temporal drift-and-return behavior emerges, which provides a quantitative signature of the QBE in the time domain. 

The temporal dynamics of the system depends strongly on the disorder strength. As $\gamma$ increases, the maximum displacement of the temporal center of mass is progressively suppressed, while its return toward the initial temporal position gradually becomes faster. In the strong-disorder regime, the temporal center of mass remains very close to its initial value throughout the evolution, indicating that temporal transport is strongly inhibited. Correspondingly, the temporal spreading of the wave packet is also suppressed at stronger disorder, consistent with a reduced localization length in the temporal lattice. Therefore, temporal disorder not only induces the boomerang return, but also stabilizes the long-time mean arrival time of the particle. These results reveal a counterintuitive role of disorder in the time domain: rather than simply destroying coherent dynamics, temporal disorder can help suppress long-time temporal drift and restore the mean arrival time toward its initial value. In this sense, the temporal QBE may provide a mechanism for disorder-assisted temporal stabilization by suppressing long-time temporal drift and preserving temporal information in periodically driven systems.

\begin{figure}[ht!]
\centering
\includegraphics[width=\linewidth]{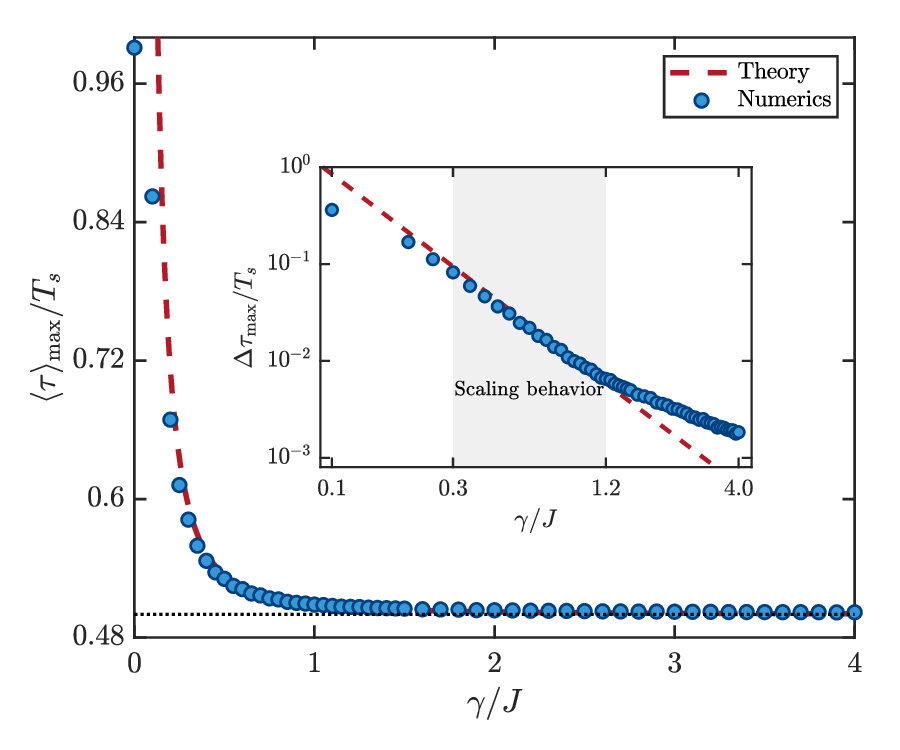} 
\caption{\corr{Maximum temporal shift $\Delta\tau_{\max}/T_s$ [Eq.~(\ref{tau_shift})] as a function of the temporal disorder strength $\gamma/J$. The temporal shift is normalized by the resonant period $T_s$.} Blue circles represent numerical simulation results, while the red dashed line indicates the theoretical scaling, $\Delta\tau_{\max} \propto (J/\gamma)^2$, from Refs.~\cite{PhysRevA.99.023629,PhysRevA.103.063316}. The inset displays the same data on a log-log scale, demonstrating the algebraic scaling within the range highlighted by the gray shaded area. Numerical data were obtained by evolving the particle for $400T_s$, with results averaged over 500 realizations of temporal disorder. All other parameters are the same as in Fig.~\ref{fig:density_W1}.}
\label{fig:TauMax_PhaseDiagram}
\end{figure}

To further quantify how the disorder strength controls the amplitude of the drift of the temporal QBE, we define the maximum temporal shift as
\begin{equation}
\label{tau_shift}
    \Delta\tau_{\max}=\langle\tau\rangle_{\max}-\langle\tau\rangle_{\mathrm{init}},
\end{equation} 
where $\langle\tau\rangle_{\text{init}}$ and $\langle\tau\rangle_{\max}$ denote the initial and maximal average temporal locations, respectively. As shown in Fig.~\ref{fig:TauMax_PhaseDiagram}, $\Delta\tau_{\max}$ decreases rapidly with increasing disorder strength. More importantly, in the intermediate disorder window $0.3\leq \gamma/J \leq 1.2$, the numerical results exhibit a clear power-law scaling, $\Delta\tau_{\max}\propto (J/\gamma)^2$, which is consistent with the effective one-dimensional Anderson localization picture \cite{PhysRevA.99.023629,PhysRevA.103.063316}. In the weak-disorder regime, the characteristic transport length is known to scale as $\gamma^{-2}$. Therefore, the observed suppression of the maximum temporal shift can be understood as a consequence of the reduction of the effective transport length in the temporal lattice. The deviations from this power-law scaling at the two ends of the disorder range can also be understood physically. In the very weak-disorder regime, the characteristic transport length becomes so large that a finite size of the lattice starts to limit the wave-packet spreading before the expected scaling behavior is fully developed. In contrast, in the strong-disorder regime, multiple scattering becomes dominant and the weak-disorder perturbative picture is no longer quantitatively accurate. These results provide further support for interpreting the temporal QBE as a dynamical consequence of temporal Anderson localization.

\corr{It is worth comparing the conditions necessary for observing Anderson localization in the kicked rotor model \cite{PhysRevX.12.011035} with those in our system. While both the kicked rotor and the system considered in this manuscript belong to the class of Floquet systems, the demonstration of Anderson localization and, consequently, the QBE requires completely different regimes. In the case of the kicked rotor, Anderson localization emerges in the chaotic regime, where the classical phase space is dominated by a chaotic sea. In that case, Anderson localization can be observed in momentum space, where quantum effects suppress classical diffusion.} 

\corr{In the system considered in the present paper, we deal with a weak resonant perturbation that creates resonant islands in the classical phase space but is too weak to destroy them. If we add an additional, very weak fluctuating perturbation, the islands in the classical description differ slightly from each other, but nothing dramatic happens. In the quantum description, however, we can reduce the description to a tight-binding model with disorder, and Anderson localization emerges. Although the particle is described by a standard tight-binding model with disorder, the basis in which this model is defined is not a static basis of Wannier states, but rather a basis of localized wave packets that evolve periodically along the resonant trajectory. This latter fact indicates that we are dealing with a Floquet system.}

\section{Conclusion} \label{sec:discussion}

In summary, we have shown that the quantum boomerang effect can also be observed in the time domain. To demonstrate this, we have considered a particle moving in a box which is periodically shaken. If the shaking is resonant with the particle motion, the system reveals a time-crystalline structure which can be effectively described by means of a tight-binding model. In the presence of temporal disorder, we observe that the particle exhibits a clear drift-and-return behavior in time, which serves as the temporal analogue of the conventional QBE in disordered spatial lattices.

\corr{To understand the physics behind this phenomenon intuitively, it is instructive to look at the underlying dynamics. A classical resonant trajectory corresponds to the evolution of a particle with a period of exactly $T_s$. If we choose the initial conditions close to this trajectory, we generally observe a non-periodic evolution that nevertheless occurs around the periodic resonant motion. In the quantum description, an initially localized wave packet will spread during the evolution unless temporal fluctuations of the driving force are present. These fluctuations, much like disorder in standard Anderson localization, lead to destructive interference and the suppression of the wave packet spreading, which is captured by reducing the description of the system to the effective Hamiltonian \eqref{tb_full}. Consequently, if the wave packet is initially given a non-zero velocity relative to a classical particle moving along the resonant trajectory, it initially moves away from that particle. However, due to the presence of temporal fluctuations of the driving force, the wave packet subsequently turns back toward the particle it escaped from. This drift-and-return mechanism constitutes the core of the temporal analog of the spatial QBE presented in this manuscript.}

We found that temporal disorder plays a crucial and counterintuitive role in \corr{the temporal QBE.} Rather than simply destroying coherent dynamics, disorder suppresses long-time temporal drift, accelerates the boomerang return, and stabilizes the long-time mean arrival time of the particle. In the intermediate disorder regime, the maximum temporal shift follows a clear power-law scaling, $\Delta\tau_{\max}\propto (J/\gamma)^2$, consistent with the effective one-dimensional Anderson-localized picture. Our results establish temporal Anderson localization as the physical origin of the temporal QBE and suggest a possible route toward disorder-assisted temporal stabilization in periodically driven quantum systems.

\begin{acknowledgments}
This research was funded by the National Science Centre, Poland, Project No. 2021/42/A/ST2/00017 \corr{and in part supported by PL-Grid Infrastructure, Project No. PLG/2025/018837.}
\end{acknowledgments}

\appendix
\renewcommand{\appendixname}{APPENDIX}

\section{Classical Secular Hamiltonian}
\renewcommand{\theequation}{A\arabic{equation}} 
\setcounter{equation}{0}

\corr{We start from the classical Hamiltonian of a particle of unit mass in a one-dimensional box of length $L=\pi$, subject to a periodic shaking force:}
\begin{equation}
    H = \frac{p^2}{2} + \lambda x \cos(\omega t).
\end{equation}
\corr{For the unperturbed system ($\lambda=0$), the particle undergoes free motion with elastic collisions at the boundaries. By introducing the action-angle variables $(I, \theta)$ \cite{Lichtenberg1992}, we have $I = |p|$ and $\theta = \Omega t$, with $\Omega = I$ is the frequency of the unperturbed motion for the chosen energy. The position $x$ satisfies $x(\theta) = |\theta|$ for $\theta \in [-\pi, \pi]$. Therefore one can expand $x(\theta)$ in a Fourier series as $x(\theta) = \sum_{n=-\infty}^{\infty} h_n e^{in\theta}$, where $h_0 = \pi/2$, and for $n \neq 0$:}
\begin{equation}
    h_n = \frac{1}{2\pi} \int_{-\pi}^{\pi} |\theta| e^{-in\theta} d\theta = \frac{(-1)^n - 1}{\pi n^2},
\end{equation}
\corr{resulting in $h_n=-2/(\pi n^2)$ for odd $n$, and $h_n=0$ for even $n$.}

\corr{To describe the dynamics near the $s$-th resonance, $\omega=s\Omega$ where $s$ is integer, we switch to the frame moving with the resonant trajectory via the canonical transformation, yielding $\tilde{I} = I$ and $\tilde{\theta} = \theta - \frac{\omega}{s}t$. Then the Hamiltonian in such a rotating frame is:}
\begin{equation}
\begin{split}
    \tilde{H} &= H + \frac{\partial F_2}{\partial t} \\
    &= \frac{\tilde{I}^2}{2} - \frac{\omega}{s}\tilde{I} + \lambda \sum_{\mathclap{n=-\infty}}^{\infty} h_n e^{in\left(\tilde{\theta} + \frac{\omega}{s}t\right)} \cos(\omega t),
\end{split}
\end{equation}
\corr{where $F_2$ is the generating function of the canonical transformation to the moving frame.}

\corr{In the moving frame the particle evolves slowly if $\tilde I\approx I_s=\omega/s$ and $\lambda$ is small, i.e.}
\begin{equation}
\begin{aligned}
    \dot{\tilde \theta} &= \tilde I-\frac{\omega}{s}+{\cal{O}}(\lambda)\approx {\cal{O}}(\lambda), \\
    \dot{\tilde I} &= {\cal{O}}(\lambda).
\end{aligned}
\end{equation}
\corr{Then, we may apply the secular approximation approach \cite{Lichtenberg1992}, i.e. we may average the Hamiltonian over time while keeping $\tilde I$ and $\tilde \theta$ fixed. Averaging over the resonant period $T_s=sT=2\pi s/\omega$, retains only the time-independent terms in the rotating frame. Such secular terms correspond to $n=\pm s$ in the sum in Eq.~(A3). Since $h_s=h_{-s}$, the secular Hamiltonian becomes:}
\begin{equation}
    \begin{split}
        H_{0,eff}^{C} &= \frac{\tilde{I}^2}{2} - \frac{\omega}{s}\tilde{I} + \lambda h_s \cos(s\tilde{\theta}) \\
        &=\frac12 \left(\tilde I-\frac{\omega}{s}\right)^2+ \lambda h_s \cos(s\tilde{\theta})-\frac{\omega^2}{s^2}.
    \end{split}
\end{equation}
\corr{Omitting the last constant term and defining}
\begin{equation}
    P=\tilde I-I_s ,
\end{equation}
\corr{we obtain the final form of the classical effective Hamiltonian}
\begin{equation}
    H_{0,\mathrm{eff}}^C \approx \frac{P^2}{2} + \tilde V\cos(s\tilde\theta), \qquad \tilde V=-\frac{2\lambda}{\pi s^2}.
\end{equation}
\corr{This arrives at the classical secular effective Hamiltonian given in Eq.~(4) of the main text.}

\section{Quantum Secular Hamiltonian}
\renewcommand{\theequation}{B\arabic{equation}} 
\setcounter{equation}{0}

\corr{In the quantum description, we also start with the exact time-dependent Hamiltonian of the particle moving in the box potential, see Eq.~(1). Solving the eigenvalue problem of the unperturbed Hamiltonian $H_0 = p^2/2$, we obtain the eigenvalues $E_n = n^2/2$ and the corresponding eigenstates, which we denote by $|n\rangle$. Similarly to the classical case, we first transform to the moving frame by means of the unitary transformation}
\begin{equation}
    U(t) = \exp(-i\hat{n} \omega t/s),
\end{equation}
\corr{where $\hat{n}|n\rangle = n|n\rangle$. In the moving frame, the Hamiltonian reads}
\begin{equation}
    \tilde{H} = U^\dagger H U - i U^\dagger \partial_t U = H_0 - \frac{\omega}{s}\hat{n} + \lambda \cos(\omega t) U^\dagger x U.
\end{equation}
\corr{Let us calculate the matrix elements of this Hamiltonian in the unperturbed eigenbasis,}
\begin{equation}
    \begin{split}
        \langle n'|\tilde{H}|n\rangle &= \left(\frac{n^2}{2} - \frac{\omega}{s}n\right)\delta_{n',n} \\
        &\quad + \lambda \langle n'| x |n\rangle e^{i\omega(n'-n)t/s} \cos(\omega t).
    \end{split}
\end{equation}
\corr{We are interested in the resonant behavior, meaning that the states of interest can be described by superpositions of the unperturbed eigenstates $|n\rangle$ with $n$ close to the resonant quantum number $n_s \approx \omega/s$. For such states, the group velocity of the unperturbed motion is very small, i.e.,}
\begin{equation}
    \frac{d}{dn}\left(\frac{n^2}{2} - \frac{\omega}{s}n\right) = n - \frac{\omega}{s} \approx 0,
\end{equation}
\corr{and for a weak perturbation, the evolution of the system is very slow. This allows us to average the matrix elements over time, which yields the matrix elements of the quantum effective Hamiltonian:}
\begin{equation}
\begin{aligned}
\langle n' | H_{0,\mathrm{eff}}^{Q} | n \rangle
&= \int_0^{T_s} \frac{dt}{T_s}
   \langle n' | \tilde{H} | n \rangle  \\
&= \left( \frac{n^2}{2} - \frac{\omega}{s} n \right)
   \delta_{n',n}  \\
&\quad - \frac{\lambda}{\pi}
   \left[
   \frac{1}{s^2}
   - \frac{1}{(2n+s)^2}
   \right]
   \delta_{n',n+s}  \\
&\quad - \frac{\lambda}{\pi}
   \left[
   \frac{1}{s^2}
   - \frac{1}{(2n-s)^2}
   \right]
   \delta_{n',n-s}.
\end{aligned}
\end{equation}
\corr{where we used the matrix elements of the position operator,}
\begin{equation}
    \langle n'|x|n\rangle = \frac{1}{\pi} \left[ \frac{(-1)^{n'-n} - 1}{(n'-n)^2} - \frac{(-1)^{n'+n} - 1}{(n'+n)^2} \right], 
\end{equation}
\corr{with the unperturbed eigenstates given by $\langle x|n\rangle = \sqrt{2/\pi}\sin(nx)$.}

\bibliographystyle{apsrev4-2}   
\bibliography{refs}             

\end{document}